\documentclass{article}
\usepackage{graphicx,amsmath}

\def\giorno{17 March 2008}

\newcommand{\beq}{\begin{equation}}
\newcommand{\eeq}{\end{equation}}
\newcommand{\feq}{\end{equation}}
\newcommand{\lb}{\label}
\newcommand{\bea}{\begin{equationarray}}
\newcommand{\eea}{\begin{endequationarray}}
\def\a{\alpha}
\def\b{\beta}
\def\ga{\gamma}
\def\de{\delta}
\def\la{\lambda}
\def\k{\kappa}
\def\s{\sigma}
\def\th{\theta}
\def\vphi{\varphi}
\def\vth{\vartheta}
\def\eps{\varepsilon}
\def\rd{\dot{r}}
\def\vphid{\dot{\vphi}}
\def\vthd{\dot{\vth}}

\def\R{\mathcal{R}}

\def\({\left(}
\def\){\right)}
\def\[{\left[}
\def\]{\right]}

\def\L{{\mathcal L}}
\def\unm{\frac{1}{2}}

\def\^#1{\widehat{#1}}

\begin{document}

\title{Twist solitons in complex macromolecules: from DNA to polyethylene}

\author{Mariano Cadoni\footnote{{\tt mariano.cadoni@ca.infn.it}},
Roberto De Leo\footnote{{\tt roberto.deleo@ca.infn.it}}, Sergio
Demelio\footnote{{\tt sergio.demelio@ca.infn.it}} \\
{\it  Dipartimento di Fisica, Universit\`a di Cagliari} \\
and {\it I.N.F.N., Sezione di Cagliari,} \\
{\it Cittadella Universitaria, 09042 Monserrato (Italy)} \\
 { } \\
Giuseppe Gaeta\footnote{{\tt
gaeta@mat.unimi.it}} \\
{\it Dipartimento di Matematica, Universit\`a di
Milano,} \\
{\it via Saldini 50, 20133 Milano (Italy)} }

\date{\giorno}

\maketitle

\begin{abstract}\noindent
DNA torsion dynamics is essential in the transcription process;
simple models for it have been proposed by several authors, in
particular Yakushevich (Y model). These are strongly related to
models of DNA separation dynamics such as the one first proposed
by Peyrard and Bishop (and developed by Dauxois, Barbi, Cocco and
Monasson among others), but support topological solitons. We
recently developed a ``composite'' version of the Y model, in
which the sugar-phosphate group and the base are described by
separate degrees of freedom. This at the same time fits
experimental data better than the simple Y model, and shows
dynamical phenomena, which are of interest beyond DNA dynamics. Of
particular relevance  are  the mechanism for selecting the speed
of solitons by tuning the physical parameters of the non linear
medium and the hierarchal separation of the relevant degrees of
freedom in ``master'' and ``slave''. These mechanisms apply not
only do DNA, but also to more general macromolecules, as we show
concretely by considering polyethylene.
\end{abstract}

\vfill\eject

\section*{Introduction}

Following the early works of Davydov on solitons in biological
systems \cite{Dav}, it has been conjectured since a long time
\cite{Eng} that nonlinear excitations -- in particular, kink
solitons or breathers -- could be present in the DNA double chain
and could play a functional role, in particular in the processes
of DNA denaturation and transcription.

This general idea meets of course essential difficulties when one
tries to translate it into quantitative terms due to the
formidable complexity of the DNA molecule \cite{CD,Sae}. This is
organized in two helices; each of them is composed of adjoining
nucleotides. A nucleotide consists of a unit of the
sugar-phosphate backbone (identical in each nucleotide) and an
attached nitrogen base (this can be of four different types; the
base at a given site on one chain uniquely determines the base at
the same site on the other chain, as they must be one of the
Watson-Crick pairs). This makes 30-35 atoms and hence about 100
classical degrees of freedom for each nucleotide; each helix is
then made, depending on the species, of $10^6 - 10^9$ nucleotides.

In view of the quasi-regular structure of DNA -- and despite the
fact genetic information is embodied in the non-regular part of
the structure -- it is quite reasonable to start the modelling by
considering a polymer made of identical units (the nucleotides),
deferring taking into account the actual base sequence and hence
inhomogeneities in the structure to a later moment (and to
computer simulations rather than analytical investigation).

Needless to say, DNA like any other molecule actually obeys
quantum rather than classical mechanics. The first consequence of
this is that nucleotides can be realistically thought as made of
rather rigid subunits, and one can just consider the degrees of
freedom of these subunits \cite{CD,GRPD,Sae}. Under closer
scrutiny, it turns out that some of these degrees of freedom are
more easily excited and hence dominant, i.e. those related to a
radial movement of the bases away from the double helix axis, and
those related to rotations of the bases and the sugar ring in a
plane nearly orthogonal to the double helix axis; in the standard
nomenclature of DNA deformations \cite{CD,Sae,GRPD}, they
correspond respectively to {\it stretch } and {\it opening}.

These considerations are at the basis of DNA modelling as
considered in the Nonlinear Mathematics and Theoretical Physics
communities, where one aims at reproducing significant
experimental observations on the basis of models with few degrees
of freedom per nucleotide. These cannot substitute for more
massive quantum chemistry computations, but could identify
relevant degrees of freedom, hence help in organizing our
understanding of the complex DNA dynamics. It is worth stressing,
in this respect, that DNA is not only complex structurally, but
also performs a great wealth of biological tasks. It is thus not
impossible that one can consider different models of it depending
on the biological process one aims at modelling; from this
perspective, considering only a few degrees of freedom per
nucleotide {\it in a model aiming at a specific mode of DNA
dynamics, relevant in a specific process}, is quite reasonable
despite the underlying overall complexity of the molecule and its
spatial organization.\footnote{The models we will consider take
into account only the double helical structure, disregarding the
way this double helix is organized in three-dimensional space;
that is, we are actually focusing at the DNA structure on small
length scales.}

\begin{figure}
  \centerline{\includegraphics[width=250pt]{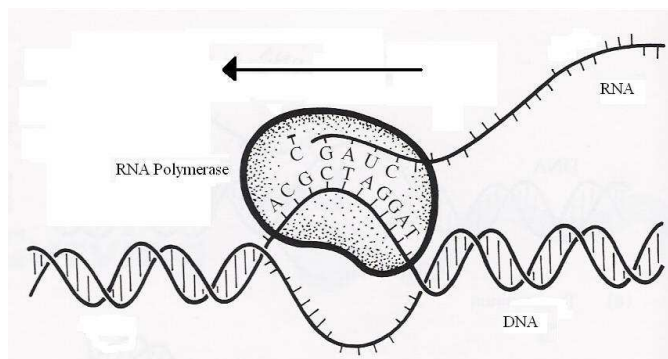}}
  \caption{A schematic view of the transcription process. RNA
Polymerase reads the bases sequence and produces RNA messenger;
the reading requires local unwinding of DNA double helix in a
``transcription bubble'' region involving about 20 base sites.
 When active in transcription, RNA Polymerase travels along DNA at a
speed of about $0.5 - 1.0 \times 10^3$ bases per second, and the
transcription bubble moves along the molecule at the same speed.
Adapted with modifications from \cite{CD}.} \label{fig:trascri}
\end{figure}

As mentioned above, simple DNA models are primarily focusing on
DNA denaturation and transcription; more specifically, they aim at
describing the deformation of DNA structure corresponding to the
two main degrees of freedom mentioned above, ``radial'' ones
(movement of the bases directly away from the double helix axis),
thought to be relevant in DNA denaturation; and ``torsional'' ones
(rotations of the bases and the sugar ring in a plane orthogonal
to the double helix axis), thought to be relevant in DNA
transcription where local untwisting of the double helix would
make possible to RNA Polymerase to access the base sequence
without disrupting the double helix
\cite{CD,Eng,GRPD,Sae,YakuBook}; see Figure \ref{fig:trascri}.

In this note we will first discuss these models, in particular
some concrete models widely studied in the literature, with their
success and limitations; and then consider a specific concrete
``composite'' model we recently proposed for the torsional
dynamics and which on the one hand is free of limitations
pertaining to other torsional models, and on the other hand shows
phenomena which are obviously not specific to it, and could be of
interest -- theoretical but also applicative -- in a much wider
nonlinear mechanics context. In order to illustrate this wider
applicability, in section \ref{sec:poly} we also discuss how these
phenomena show up in a different macromolecule, i.e.
polyethylene.\footnote{Our discussion of the DNA composite model
will go over our recent results \cite{CDG1,CDG2,CDG3,CDG4}; on the
other hand, the discussion of how the general mechanisms devised
in that context apply to polyethylene -- based on the
Zhang-Collins model \cite{ZhC} -- is new.}

We will not discuss DNA structure and its Physics, for which the
reader is referred e.g. to \cite{CD,Sae} and \cite{FK2,Vol}
respectively. Experimental studies of single-molecule DNA dynamics
are discussed e.g. in \cite{Lavery,Ritort,Strick}; for general
discussion of modelling DNA at different scales, see \cite{BFLG}.

\section{Mechanical models of DNA}
\label{sec:mechmod}

A number of mechanical models of the DNA double chain have been
proposed over the years, focusing on different aspects of the DNA
molecule and on different biological, physical and chemical
processes in which DNA is involved. A discussion of such attempts
is given in the books by Yakushevich \cite{YakuBook} and by
Dauxois and Peyrard \cite{PD}, as well as in the review paper by
Peyrard \cite{PeyNLN}.

In recent years, two models have been extensively studied in the
Nonlinear Physics literature. These are the ``radial'' model by
Peyrard and Bishop \cite{PB} (and the extensions of this
formulated by Dauxois \cite{DauPLA}, Dauxois, Peyrard and Bishop
\cite{PBD}, and later on by Barbi, Cocco, Peyrard and Ruffo
\cite{BCP,BCPR}; see also Cocco and Monasson \cite{CM}. More
recent advances are discussed in \cite{PeyNLN} and references
therein); and the ``torsional'' one by Yakushevich \cite{YakPLA},
which had precursors discussed in \cite{YakuBook} and is put in
perspective within a hierarchy of DNA models in
\cite{YakPhysD,YakJBS}. We will refer to these as the PB and the Y
models respectively.\footnote{Here we will discuss, for the sake
of brevity, only the ``planar'' versions of these models, i.e.
overlook so called ``helicoidal'' interactions. These are
interactions between bases which are not first-neighboring in the
chain sequence but which come to be near in three-dimensional
space due to the helical geometry of the DNA molecule. Considering
these introduce qualitative differences in the dispersion
relations, both in the PB \cite{DauPLA} and in the Y model
\cite{GaePLA}; see \cite{GaeJBP,GRPD} for a discussion. The same
will apply for the composite Y model discussed in
Sect.\ref{sec:compYmod}; see \cite{CDG1} for its full version,
taking into account ``helicoidal'' interactions.}

The interplay between radial and torsional degrees of freedom of
bases is considered organically in the Barbi-Cocco-Peyrard (BCP)
model \cite{BCP,BCPR,CM}; the latter was formulated as an
extension of the Peyrard-Bishop-Dauxois model \cite{PB,PBD}, i.e.
in the context of ``radial'' dynamics.

\subsection{The PB model}
\label{subsec:PBmod}

In PB-like models, the bases can only move radially away from the
double helix axis. The potential energy corresponds to stacking
interactions between successive bases on each chain, and pairing
interactions between bases at corresponding sites on opposite
chains \cite{CD,GRPD,PD,Sae}.

We denote by $r_n^{(\pm)} \in \R_+$ the position of the base at
site $n$ on the helix ${(\pm)}$.  It is convenient, for our
present purposes, to limit discussion of this and other models to
symmetric configurations, i.e. configurations with $r_n^{(+)} (t)
= r_n^{(-)} (t) = r_n (t)$ at all times. In this case, the
Lagrangian describing the PB model is (with $m$ the mass of bases)
\beq\label{eq:lagPB} \L_{{\rm PB}} \ = \ \sum_n m (\rd_n)^2 \ - \
K_s \, \(r_{n+1} - r_n \)^2 \ - \ V_p (r_n) \ . \eeq In the PB
model, $V_p$ is a Morse potential with a minimum at the
equilibrium position, corresponding to $r_n = \rho \simeq 2 \AA$:
\beq\label{eq:morse} V_p (r) \ = \ D \, \( \exp \[ - \a (r - \rho)
\] - 1 \)^2 \ . \eeq

The Euler-Lagrange (EL) equations for $\L_{{\rm PB}}$ are of
course
 \beq\label{eq:ELPBdisc} m \ddot{r}_n \ = \ \, K_s \, (r_{n+1}
- 2 r_n + r_{n-1} ) \ - \ \unm {V_p}' (r_n) \ . \eeq We will pass
to consider the continuum approximation for these, i.e. substitute
the infinite array of scalar variables $\{ r_n (t) \}$ with the
interpolating field $R (x,t)$, such that $R(n \de ,t ) \approx r_n
(t)$; here $\de \simeq 3.4 \AA$ is the distance between successive
base pairs positions along the double helix axis. Using now a
second order approximation, \beq R (x \pm \de , t) \approx R(x,t)
\pm \de R_x (x,t) + (\de^2/2) R_{xx} (x,t) \ , \eeq and writing
for short $\k = K_s \de^2$, the \eqref{eq:ELPBdisc} yield the {\it
nonlinear wave equation}\footnote{The same equation is also
obtained passing to the continuum approximation directly in the
Lagrangian, i.e. considering the Lagrangian density $L_{{\rm PB}}
= R_t^2 - K_s R_x^2 - V_p (R) $. The same holds for the other
models considered below.}
 \beq\label{eq:ELPB} m R_{tt} \ = \ \k \, R_{xx}
\ - \ (1/2) [d V(R)/d R] \ . \eeq

It can be shown that this supports {\it breathers}; their size and
oscillation frequency (choosing parameters so that $V_p$ describes
as far as possible a Hydrogen-bond interaction) are compatible
with those observed in real DNA. The discrete model can be put in
a thermal bath and numerically simulated, and again one observes a
behavior compatible with the one observed in DNA denaturation
\cite{PeyNLN,PD}. More refined versions of this model \cite{PBD}
consider modified expressions for the stacking energy, improving
-- also qualitatively -- the correspondence with actual DNA
behavior; we refer again to \cite{PeyNLN,PD} for details.

\subsection{The Y model}
\label{subsec:Ymod}

In the simple models for DNA torsional dynamics, one studies a
system of nonlinear equations which in the continuum limit reduce
to a pair of sine-Gordon (SG) type equations; the relevant
nonlinear excitations are kink solitons -- which are solitons in
both dynamical and topological sense -- which describe the
unwinding of the double helix in a ``bubble''.

The main biological interest of these model lies in the
identification of this unwound bubble with the transcription
region (this is indeed an ``open bubble'' of about 20 bases, to
which RNA Polymerase (RNAP) binds; the RNAP travels along the DNA
double chain, and so does the unwound region). The idea of
Englander et al. \cite{Eng} was that the open bubble could
correspond to nonlinear excitations and thus be present due to the
nonlinear dynamics of the DNA double helix itself; the RNAP would
then use them to travel along DNA. In this way a number of
questions -- in particular, concerning energy flows -- would
receive a simple explanation. Note that their model, and
subsequent ones continuing their research, are not concerned with
the DNA-RNAP complex, but the dynamics of the DNA double helix
alone.

To be specific, let us consider the model proposed by Yakushevich
(see e.g. \cite{YakuBook} for similar models proposed earlier on
by other authors, starting with \cite{Fedyanin,Takeno,Yomosa});
now the degrees of freedom for the rotation of the base at site
$n$ on the chain $(\pm)$ will be denoted as $\vphi_n^{(\pm)}$, and
again we restrict to the symmetric case, i.e. enforce
$\vphi_n^{(+)} (t) = \vphi_n^{(-)} (t) = \vphi_n (t)$ at all times
(this makes that in the continuum approximation we will get a
single equation of SG type rather than two coupled ones). The
Lagrangian describing the Y model is (here $m$ is a moment of
inertia) \beq\label{eq:lagY} \L_{{\rm Y}} \ = \ \sum_n m
(\vphid_n)^2 \ - \ K_s \, \(\vphi_{n+1} - \vphi_n \)^2 \ - \ \^V_p
(\vphi_n) \ ; \eeq the choice by Yakushevich for the intrapair
potential was that of a simple harmonic potential\footnote{The Y
model also sets to zero the equilibrium length for this harmonic
interaction (contact approximation); this results, as observed by
Gonzalez and Martin-Landrove \cite{Gonz}, in a degeneration of the
model (it is thanks to this that we obtain exactly the SG
equation). If we go beyond the contact approximation, the
equations we obtain are more complex, dispersion relations change
quantitatively and qualitatively, but soliton solutions are little
affected by this \cite{GaeY1}; see also Figure \ref{fig:Ycompare}
below.} \cite{YakPLA}, resulting in (we stress here $r$ is a
geometrical constant, not a dynamical variable!) \beq \^V_p (\vphi
) \ = \ - 4 K_p r^2 \, \cos (\vphi ) \ . \eeq The EL equations for
$\L_{{\rm Y}}$ are a set of sine-Gordon coupled equation, \beq
\label{eq:ELYd} m \ddot{\vphi}_n \ = \ K_s \, (\vphi_{n+1} - 2
\vphi_n + \vphi_{n-1} ) \ - \ (1/2) \, {\^V_p}' (\vphi_n) \ . \eeq
Passing to continuum approximation with interpolating field $\Phi
(x,t)$ (where $\Phi n \de , t) \approx \vphi_n (t)$, like above)
and second order approximation \beq \Phi (x \pm \de , t) \approx
\Phi(x,t) \pm \de \Phi_x (x,t) + (\de^2/2) \Phi_{xx} (x,t) \ ,
\eeq the \eqref{eq:ELYd} reduce to a sine-Gordon equation \beq
\label{eq:ELY} m \Phi_{tt} \ = \ \k \Phi_{xx} \ - \ \la \, \sin
(\Phi) \ ; \eeq here of course we have made use of the explicit
form of $\^V_p$, set $\la = 2 K_p r^2$, and defined $\k$ as above.

As well known, the sine-Gordon equation supports topological
soliton solutions \cite{DNF}; these solutions are also solitons in
the dynamical sense \cite{CalDeg}. The basic soliton solution with
speed $v$ is \beq \label{eq:Ysol} \Phi (x,t) \ = \ 4 \, {\rm
arctan}
\[ \exp (\b (x - v t) )
\] \ , \eeq where we have written
\beq \label{eq:betamu} \b = \sqrt{ \frac{2 K_p r^2}{- \mu } } \ ,
\ \ \mu =  m^2 v^2 - K_s \de^2 \ . \eeq Note that $v$ is a free
parameter, subject to the condition\footnote{The existence of a
limit speed for travelling waves (and not just soliton solutions)
is related to the Lorenz invariance of the SG equation; see also
\cite{CDG3} and Sect.\ref{sec:speed} below.} $|v| < v_* =
\sqrt{K_s} (\de / m)$. A selection of the speed based on energy is
also not present, as the soliton energy has a very weak dependence
on its speed except for $v \simeq \pm v_*$, see \cite{GaeSpeed}.

It has been shown that the Y model gives a correct prediction of
quantities related to small amplitude dynamics, such as the
frequency of small torsional oscillations; and also of quantities
related to fully nonlinear dynamics, such as the size of solitonic
excitations describing transcription bubbles \cite{GRPD,YakuBook}.

On the other hand, the Y model is not capable of providing a
satisfactory prediction for other quantities: in particular, if we
try to fit the observed speed of transversal waves along the chain
\cite{YakuBook}, this is possible only upon assuming unphysical
values for the coupling constants \cite{YakPRE}. That is, with a
physical value of the parameters -- in particular, for $K_s
\approx 120 {\rm KJ/mol}$ -- the Y model predicts\footnote{It
follows from the discussion in \cite{YakPRE} that $v = A
\sqrt{K_s}$ with $A \approx 28.3335$; see in particular the
non-numbered formula before eq.(7) there.} a speed $v \approx 320
{\rm m/s}$, while in order to get a speed of the order of 2 Km/s
as observed in experiments\footnote{More precisely, measures on
DNA fibers in the B-DNA conformation give $v = 1.9 {\rm Km/s}$
\cite{Hak}, while measures in DNA crystals yield a speed of 2.45
Km/s, which can grow up to 4.15 Km/s depending on counterions
concentration and chemical nature \cite{Wei}. Note only transverse
waves speed matters here, as the model does not allow for
longitudinal waves.} and \cite{Hak,YakPRE} one has to take $K_s
\approx 6000$.

Finally, we stress once again that the Y model, like the PB one
and unlike the BCP one, assumes that there is a single (angular in
this case) degree of freedom for each nucleotide.

\subsection{The BCP model}
\label{subsec:BCPmod}

In the BCP model \cite{BCP,BCPR,PeyNLN,PD}, the state of each base
is described by both a radial $r_n^{(\pm)}$ and an angular
$\vphi_n^{(\pm)}$ variable. Restricting again to symmetric
configurations, and using variables $\{r_n,\vphi_n \}$, the BCP
Lagrangian is \beq \label{eq:LagBCP} \begin{array}{rl} \L_{{\rm
BCP}} \ =&  \sum_n m (\rd_n^2 + r_n^2 \vphid_n^2) \ - \ \sum_n V_p
(r_n) \ + \\ & \ - \ \sum_n K_s \( L - \sqrt{\de^2 + r_{n+1}^2 +
r_n^2 - 2 r_{n+1} r_n \cos (\vphi_{n+1} - \vphi_n)} \)^2 \ + \\ &
\ - \ \sum_n G_0 (\vphi_{n+1} - 2 \vphi_n + \vphi_{n-1} )^2 \ ,
\end{array} \eeq with $V_p$ the same Morse potential as above (in the
non
symmetric case, pairing would depend on $\phi$ variables as well)
and $L$ the equilibrium distance between bases in
three-dimensional space\footnote{The geometry of the BCP model
assumes the distance $h_n$ between successive base pairs planes
(at sites $n$ and $n+1$) is constant and equal to $\de$, while the
length $\ell_n$ of the sugar-phosphate backbone unit connecting
them can vary. A similar model where $\ell_n$ is fixed and $h_n$
can vary has been formulated by Cocco and Monasson (CM model)
\cite{CM}.}. As for the last term in \eqref{eq:LagBCP}, this is a
curvature term whose role is to avoid ``zig-zag'' configurations,
possible for $G_0 = 0$; it should thus be omitted in the continuum
approximation.

In this case, proceeding as above (or more simply using the
continuum version of the Lagrangian) one obtains
\cite{GV,PeyNLN,PB} in the continuum approximation the EL
equations \beq \label{eq:ELbcp}
\begin{array}{rl} m R_{tt} \ =& m R \Phi_t^2  -
\ga (R P_x^2 - R_{xx} ) - (1/2) V'(R) \ , \\
m R^2 \Phi_{tt} \ =& - 2 m R R_t \Phi_t + \ga (R^2 \Phi_{xx} + 2 R
R_x \Phi_x ) \ . \end{array} \eeq It has been shown that for $L >
\de$ this supports nontrivial topological soliton solutions; the
determination of these reduces to the study of a single equation
thanks to an exact conservation law (the model is invariant by
global rotations), but solitons are only determined numerically
\cite{GV}.

The BCP model is quite successful in describing breather
excitations of the DNA double chain, and its predictions (or those
of the cognate CM model) fit well experimental observations
\cite{PeyNLN,PB} related to the DNA denaturation process.

\section{Composite model of DNA torsional dynamics}
\label{sec:compYmod}

A closer look at nucleotides structure and conformation
\cite{CD,Sae} shows that torsional motions can take place both as
a rotation of the nitrogen base with respect to the sugar ring and
as a rotation of the sugar-phosphate group\footnote{More
precisely, of the sugar ring around the $P-O-C-C-C-O-P...$
chain.}; it should be stressed that the former one is subject to
steric hindrances, i.e. is limited by interactions with other
parts of the DNA molecule.

Thus, one should consider ``composite'' torsional models, in which
we describe the state of each nucleotide by {\it two} independent
angular degrees of freedom, one related to the sugar-phosphate
group and one to the nitrogen base; see Figure \ref{fig:modello1}
for details. Note that in this scheme one is {\it not }
considering ``radial'' (stretch) motions.

\subsection{The model}

\begin{figure}
\includegraphics[width=300pt]{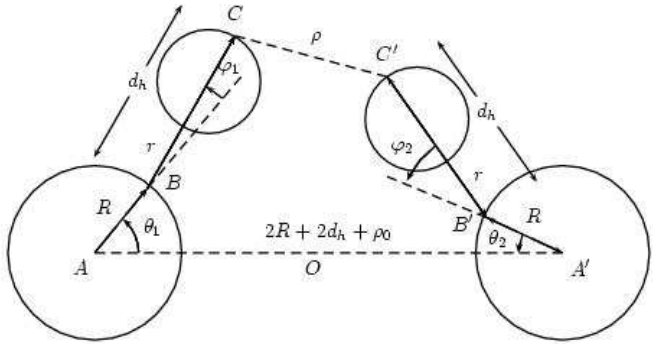}\\
  \caption{A base pair in the composite Y model (reproduced from
\cite{CDG1}).
  The origin of the coordinate system is in O; the angles $\theta_1$
  and $\theta_2$ correspond
  to torsion of sugar-phosphate backbone with respect to the
equilibrium
  B-DNA conformation; the angles $\vphi_1$ and
  $\vphi_2$ between A'B' and B'C' correspond to rotation of bases
around
  the $C - N$ bond linking them to the nucleotide. All angles are in
counterclockwise
  direction.}
  \label{fig:modello1}
\end{figure}

Such a model, described graphically in Figure \ref{fig:modello1},
was recently put forward and studied by three of us \cite{CDG1}
(see also \cite{CDG2} for a discussion of it focusing on its
mathematical features); it shows some phenomena -- to be discussed
in later sections -- which are not specific to it but apply to a
more general class of models and could be of interest in fields
quite far from DNA dynamics \cite{CDG3,CDG4}.

Our model can be considered as an extension of the Y model. It
turns out that the simple Y model captures to a large extent the
essential features of the nonlinear dynamics of the composite
model. On the other hand, the more realistic geometry of the
composite model yields a relevant improvement of the descriptive
power of the model at both the conceptual and the phenomenological
level; in particular, the composite Y model allows for a more
realistic choice of the physical parameters.

The different degrees of freedom we use will play a fundamentally
different role in the description of DNA nonlinear dynamics. The
backbone degrees of freedom (recall the rotations they describe
are not limited) are ``topological'' and play to some extent a
more relevant role, in that the solitons are mainly associated to
them; while those associated to the base (recall the associated
rotations are subject to steric hindrances) are ``non
topological'' and represent small oscillations. These different
roles are specially clear when we consider the limit in which our
model reduces to the standard Y model, in which only the
topological degrees of freedom are present.

It should be stressed that this feature is specially interesting
in connection with the possibility (discussed elsewhere \cite{DD})
to consider more realistic models, in which differences among
bases are properly considered, as perturbations of our idealized
uniform model. As the essential features of the fully nonlinear
dynamics are related only to backbone degrees of freedom, such a
perturbation should be expected to show the same kind of nonlinear
dynamics as our uniform model.

The Lagrangian defining the composite Y model will be written as
\beq \label{eq:LAGcY} \L_{{\rm cY}} \ = \ T \ - \ \( U_t \ + \ U_s
\ + \ U_p \) \ - \ U_c \eeq with $T$ the kinetic energy, $U_t$ is
the backbone torsional potential, $U_s$ the stacking potential,
and $U_p$ the pairing potential.

Moreover $U_c$ is a constraining potential which represents the
steric hindrances to the base rotations (not accompanied by
sugar-phosphate rotations); its explicit expression is to a large
extent arbitrary, provided $\vphi_n$ are {\it de facto } bound to
a small range around zero.

Explicit expressions for the potentials are rather involved
\cite{CDG1} and will be given below in \eqref{eq:LAGcYparts},
considering again symmetric configurations only for the sake of
simplicity (see \cite{CDG1,CDG2} for the general case). Note in
there $r$ and $R$ are geometrical parameters (and $m$, $I$
represent masses and inertia moments), while the dynamical
variables are $\{ \vphi_n (t) , \vth_n (t) \}$, $n \in {\bf Z}$.

Before writing these explicit expressions, we state that $U_t$ and
$U_s$ correspond to the torsional energy of the backbone and the
base stacking energy respectively, and have simple harmonic
expressions\footnote{The possible relevance of nonlinear stacking
interactions has been recently noted in \cite{SacSgu}.} in terms
of the three-dimensional coordinates of involved nucleotide
elements (albeit a rather involved expression in terms of torsion
angles); as for the pairing interaction $U_p$, this corresponds
again to a harmonic potential in three-dimensional distance
between bases in a pair. This latter choice, albeit not natural
physically (a Morse potential as in the PB or BCP models would
perhaps be more appropriate; see \cite{GaeY2} for a discussion of
different intrapair potentials within the Y model) was made in
order to ease comparison with results obtained for the simple Y
model, i.e. in order to focus attention on the improvements which
have to be ascribed to the geometry of the model. Here are the
explicit expressions for $T$ and the $U_i$: \beq
\label{eq:LAGcYparts}
\begin{array}{rl}
T \ =&\frac{1}{2} \ \sum_n
 \ \[ m \, r^2 \, \( \vphid_n \)^2 \ + \ 2 \, m \, r \,
 ( r + R \cos (\vphi_n) ) \,
\vthd_n \, \vphid_n \ + \right. \\
 & \left. + \ \( I + m (R^2 + r^2 ) + 2 m R r
 \cos (\vphi_n) \) \( \vthd_n \)^2 \] \ ; \\
U_t \ =& \ K_t \ \sum_n \, \[ 1 \, - \, \cos \( \vth_{n+1} -
\vth_n \) \] \ ; \\
U_s \ =& \ K_s \ \sum_n \ 2 \, \[ R^2 +
r^2 \ + \right. \\
& \left. - \ R^2 \cos (\vth_{n+1} - \vth_{n} ) \, - \, r^2 \cos [
(\vth_{n+1} - \vth_{n} ) +
(\vphi_{n+1} - \vphi_{n}) ] \ + \right. \\
 & \left. - \ R r \( \cos [(\vth_{n+1} - \vth_{n})
 + \vphi_{n+1} ] + \cos [(\vth_{n+1} - \vth_{n})
 - \vphi_{n} ] \) \ + \right. \\
  & \left. + \ R r \( \cos (\vphi_{n+1}) + \cos(\vphi_{n}) \) \] \
  ; \\
 U_p \ =& \ (1/2) \, K_p \ \sum_n \, (\s_n - \rho)^2 \ , \ {\rm
 with} \\
\s_n^2 \ =& \ 4 \[ a^2 + R^2 + d_{h}^2 + 2 \( R d_{h} \cos \vphi_n
- a R \cos \vth_n - a d_{h} \cos (\vphi_n  + \vth_n )\) \] \ .
\end{array} \eeq Here $m$ is the base mass, $I$ the momentum of
inertia of the disk modelling the backbone units and and
$K_{t},K_{s},K_{p}$ are, respectively the torsional, stacking and
pairing coupling constants; $\s_n$ represents the distance between
bases in a pair (between end points of double pendulums), so that
$U_p$ is a harmonic potential in the physical distance, albeit
expressed by a non-harmonic function when dealing with angular
variables.

We work in the so-called contact approximation in which the
equilibrium distance between the two basis, $\rho_{0}$ vanishes
(see Figure (\ref{fig:modello1}). In this approximation the
parameter $a$ appearing above is given by $a= R+d_{h}$. Proceeding
as in the PB, Y and BCP cases above, we obtain field equations for
the interpolating fields $\Phi (x,t)$ and $\Theta (x,t)$; these
are rather lengthy and reported explicitly in Appendix B of
\cite{CDG1}. As we are mainly interested in soliton equations, we
will introduce the {\it travelling wave ansatz} \beq
\label{eq:TWA} \Phi (x,t) = \Phi (x - v t) = \phi (z) \ , \ \
\Theta (x,t) = \Theta (x - v t) = \theta (z) \ . \eeq The
equations for travelling wave (TW) solutions are
 \beq \label{eq:TWcY}
\begin{array}{l}
\mu r^2 \, \phi'' \ + \ \mu r (r + R \cos \phi ) \, \theta'' \ = \\
\ \ = \ - 2 a K_p (a - R) \, \sin \(\phi + \theta \) + K_s \de^2 R
r \, \sin \( \phi \) \, (\theta' )^2 \ + \\
\ \ \ - R \, \sin (\phi) \, (-2 K_p (a - R) + m r v^2 (\theta' )^2
) - \frac{dU_{c}}{d\phi}\ ; \\
\mu r (r + R \cos \phi ) \, \phi'' \ + \ [ J + \mu  (R^2 + r^2 +
2 R r \cos \phi ) \, \theta'' \ = \\
\ \ = \ - 2 a K_p \( R \sin \theta + (a - R) \sin (\phi + \theta)
\) + \mu R r \, \sin (\phi) [ (\phi')^2 + 2 \phi' \theta' ] \ .
\end{array} \eeq
Here we have simplified the notation by introducing the constants
\beq \label{eq:muJ} \mu \ = \ (m v^2 - K_s \de^2 ) \ , \ \ J \ =
(I v^2 - K_t \de^2 ) \ . \eeq

It should be stressed that the ODEs \eqref{eq:TWcY} are obtained
as a reduction of the nonlinear wave PDEs for $\Phi$ and $\Theta$;
the latter should be supplemented by boundary conditions.
Requiring that the solutions go to an equilibrium for $x \to \pm
\infty$ (and thus have finite energy), these are $\Phi (\pm
\infty,t) = \Phi_x (\pm \infty,t) = \Phi_t (\pm \infty,t) = 0$,
$\Theta (\pm \infty,t) = 2 n_\pm \pi$, $\Theta_x (\pm \infty,t) =
\Theta_t (\pm \infty,t) = 0$. These entail side conditions for
$\phi (z)$ and $\theta (z)$, i.e.
 \beq\label{eq:bc}
\phi (\pm \infty) = 0 \ , \ \theta (\pm \infty) = 2 n_\pm \pi \ ;
\ \ \phi' (\pm \infty) = 0 \ , \ \theta' (\pm \infty) = 0 \ . \eeq

We also note that the constraining potential $U_c$ makes that the
fields $(\Theta,\Phi)$, which in principles take values in $S^1
\times S^1$, are actually taking values in $S^1 \times I_0$ (where
$I_0 = (-\la , \la)$ is a real interval centered in zero, with
$\la \ll \pi$); thus $\Theta$ is a topological field, while $\Phi$
is a non-topological one (that is why the boundary conditions for
$\Phi$ do not allow nontrivial multiples of $2 \pi$).

\subsection{Physical  values of the parameters}
\label{s5}

One of the nice and striking  features of our composite model is
that it supports solitonic solutions within a fully realistic
range of all the physical parameters characterizing the DNA; this
should be compared with the situation for simple torsional models
mentioned above, where unphysical coupling constants are needed to
fit some experimental data.

Let us therefore briefly discuss how the values of the parameters
appearing in Eq. (\ref{eq:LAGcYparts}) are fixed.

There are basically two types of parameters: kinematical ( the
geometrical parameters $R,r,d_{h}, a$, the  mass $m$  and the
momentum  of inertia $I$), and dynamical (the elastic coupling
constants $K_{t}, K_{s}, K_{p}$ )\footnote{Notice that the
values of the physical parameters given in this note  differ slightly
from those of Ref (\cite{CDG1}). The new values  improve the
estimates  of the parameters but do not change the qualitative
behavior of the model.}

The kinematical parameters can be evaluated by considering the
chemical structure and the geometry of the DNA molecule. The order
of magnitude of the mean value of the bases mass is $m\sim 130$
(atomic units). For the momentum of inertia of the disk we have $I
\sim 5 \times 10^{-44} \, {\rm Kg \, m}^{2}$, whereas the values
of the geometrical parameters are given in Table 1.

\begin{table}\label{table:cY1}
\begin{center}
   \begin{tabular}{|c|c|c|c|}
    \hline
    $R$ & $r$ & $d_{h}$ & $a$ \\
    \hline
    3.1~\AA  &2.7~\AA  & 4.4~\AA & 7.5~\AA  \\
    \hline
   \end{tabular}
\caption{The geometrical parameters in the composite DNA model.}
\end{center}
\end{table}

The determination of the numerical values of the coupling
constants characterizing our model is more involved. Their order
of magnitude can be estimated by considering the typical energy of
hydrogen bonds ($K_{p}$) and the experimental results for the
torsional rigidity of the DNA chain ($K_{s}$ and $K_{t}$)
\cite{CDG1}. The results are given in Table 2.

\begin{table}\label{table:cY2}
\begin{center}
   \begin{tabular}{|c|c|c|}
    \hline
    $K_t$ & $K_s$ & $K_p$  \\
    \hline
    130 KJ/mol & 16.6 N/m & 3.5 N/m  \\
    \hline
   \end{tabular}
\caption{The dynamical parameters in the composite DNA model.}
\end{center}
\end{table}

These values allow in particular to estimate the speed of torsion
waves associated to base torsion; this will be the speed of
transverse elastic waves along the double chain.

With the above values, the speed of elastic waves is estimated to
be $v_s = \de \sqrt{K_s/m} \approx 3 {\rm Km/s}$; this is of the
right order of magnitude, and well compatible with experimental
data \cite{Hak,Wei}.

Let us stress that the geometry of our composite model makes that
using natural parameters one obtains predictions that nicely fits
with the estimates of the structural properties and binding
energies of the DNA: the induced optical frequencies and phonon
speeds are of the same order of magnitude of those experimentally
observed. This does not happen in simpler models, as stressed
above when discussing the simple Y model.

\subsection{Nonlinear dynamics and soliton solutions}
\label{sec:soliton}

Equations (\ref{eq:TWcY}) are a system of two coupled, non linear
ODEs. In general it cannot be solved analytically in closed form.
One has to resort to numerical calculations in order to show that
the system admits  solutions satisfying the boundary conditions
(\ref{eq:bc}) \cite{CDG1} (see also Ref. (\cite{DD}).

We note that the relevance of the dynamical system (\ref{eq:TWcY})
goes well beyond DNA torsional dynamics: the same kind of
equations appear in more general cases.

In fact, we can as well consider (\ref{eq:TWcY}) as describing the
continuum limit of the torsional dynamics of a single molecular
chain made of a disk and a pendulum. In this case, the pairing
interaction for the DNA double chain is replaced by an external
potential: \beq V=-4r^{2} K_{p}\( \cos\theta +\cos(\vphi+\theta) -
\frac{1}{2}\cos\vphi - 3/2 \) \ , \eeq whereas the stacking and
torsional interaction generate the $x$-derivative terms.

To further simplify our model -- and the resulting explicit
formulas -- and concentrate on essential features of interest also
beyond DNA dynamics, we also set $R=r$ in (\ref {eq:TWcY}). The
resulting equations take the much simpler form
\beq\label{simple}
\begin{array}{l}
\mu  \, \phi'' \ + \ \mu  (1 +  \cos \phi) \, \theta'' \ = \\
\ \ \ \ = \ - 4  K_p  \, \sin \(\phi+ \theta \)  -\mu \, \sin \(
\phi\) \, (\theta' )^2 \ + \ 2 K_p  \, \sin (\phi) -\frac{\partial
U_{c}}{\partial \phi}
 \ ; \\
\mu  (1 +  \cos \phi ) \, \phi'' \ + \ [ (J /r^2) + 2\mu (1 +
 \cos \phi )] \, \theta'' \ = \\
\ \ \ \ = \ - 4  K_p \(  \sin \theta + \sin (\phi + \theta) \) +
\mu  \, \sin (\phi) [ (\phi')^2 + 2 \phi' \theta' ] \ ,
\end{array} \feq

The previous  form of the equations of motion will be taken as
starting point to discuss two general features of the dynamical
system. As we will see in the next two sections these features
(the existence of a mechanism to select the speed of solitonic
solutions and the slaving of the field $\phi$) represent quite
general consequences of nonlinear dynamics. We expect  they will
have a quite broad field of application in the context of
non-linear Physics and Mechanics.

Although in the general case one can find a solitonic solution of
the system (\ref{simple}) only numerically, there is a particular
case which admits analytical solutions. This is obtained by
freezing the angle $\phi$, i.e by setting $\phi=0$; note that if
we force $\phi (z)= 0$,  we are actually considering a chain of
simple pendulums, i.e. a sine-Gordon equation.

This constraint can be accommodated in our setting in a dynamical
way, by acting on the confining potential $U_c$: this should be
made stronger and stronger and the maximum angle $\phi_0$ will
become smaller and smaller.

Setting $\phi=0$ and using $(\partial U_{c}/\partial \phi) (0)=0$,
the system (\ref{simple}) is equivalent to the  equation
\beq\lb{ge1} \mu\theta''=-2K_{p}\sin\theta \ ; \feq
the compatibility condition between the two equations of the
system (\ref{simple}) is now given by \beq\lb{comp} J=0. \feq

Equation (\ref{ge1}) has to be integrated with the boundary
conditions (\ref{eq:bc}). When $\mu<0$ and $n=1$, we have the kink
\beq\lb{fe3} \theta_{0}= 4 \arctan[e^{ \b z}],\quad\quad \phi=0 \
, \feq
where we have written, as in \eqref{eq:betamu}, $\b =
\sqrt{\frac{2K_{p}}{- \mu}}$. The solution \eqref{fe3} is of
course the same as \eqref{eq:Ysol}, i.e. the solution found in the
context of the simple Y model.

\begin{figure}
 \includegraphics[width=150pt]{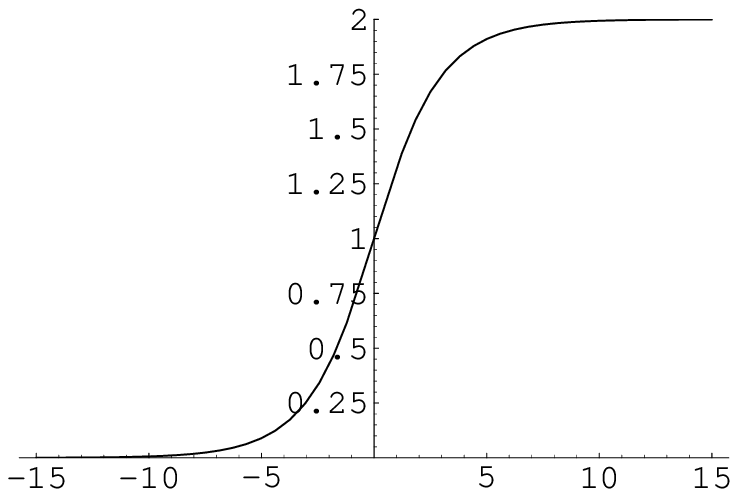} \ \
 \includegraphics[width=150pt]{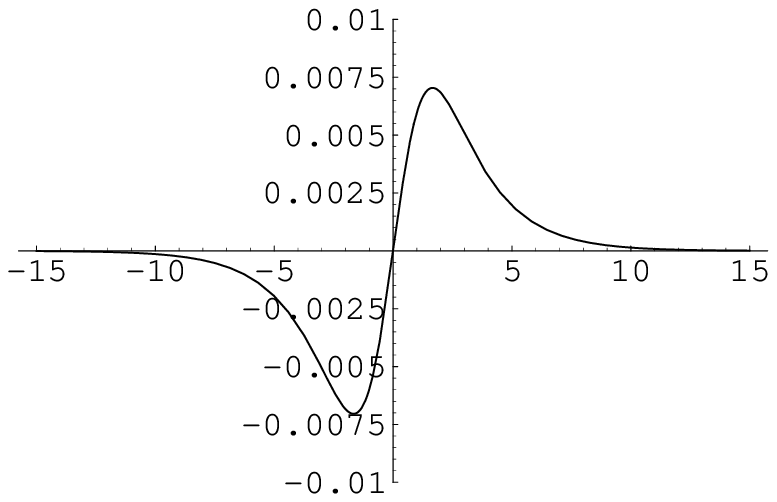} \\
\caption{The $n=1$ soliton for the composite model of Figure
\ref{fig:modello1}. We plot in thick curves the fields $\theta$
(on the left) and $\phi$ (on the right) as a function of $x$ for
the static ($v=0$) solution with physical values for the model
parameters (reproduced from \cite{CDG2}). The solution is very
similar to the corresponding one obtained for the Y model for what
concerns the $\theta$ field.} \label{fig:cYsol}
\end{figure}

As for general solutions, we note these will still be indexed by
the topological index $n$ (which refers to the $\theta$ behavior);
the full equations \eqref{simple} cannot be solved analytically in
the general case (see below for a perturbative approach), but they
can be studied and solved numerically; the solution for $n=1$ is
displayed in Figure \ref{fig:cYsol}.

\subsection{Discussion}

It is quite clear that a weak point of this model is represented
by the choice of the exceedingly simple potential $U_p$; and also
by the the simplifying assumption $\rho=0$ (see
\eqref{eq:LAGcYparts} and comments thereafter).

As mentioned above, this choice was justified by the will to ease
comparison with results obtained with the simple Y model, i.e. to
be able to focus on new features depending only on the more
articulated geometry of the model.\footnote{And also -- in the
present case -- to be able to discuss interesting phenomena
without being forced to tackle technically hard computations which
could hide the physical and mechanical meaning of the results.}

Needless to say, one should then consider the same model with more
realistic pairing potentials -- e.g. with the Morse potential used
in the PB and BCP models (this is being done \cite{DD}, and yields
quite interesting preliminary results).

\begin{figure}
  \includegraphics[width=150pt]{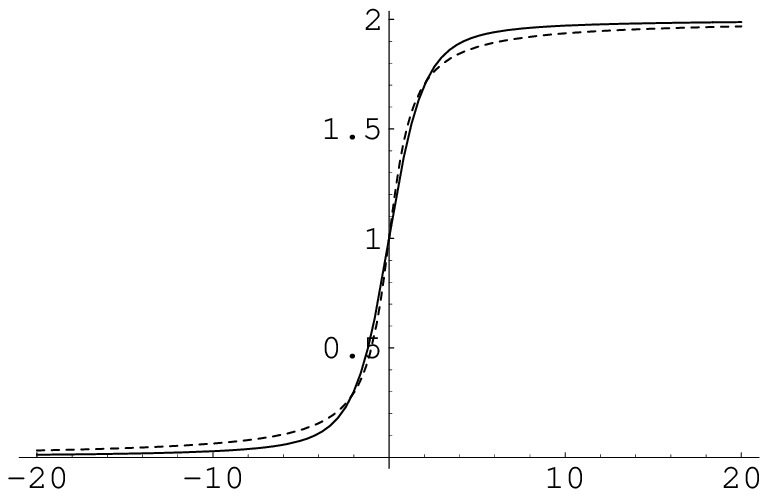} \ \
    \includegraphics[width=150pt]{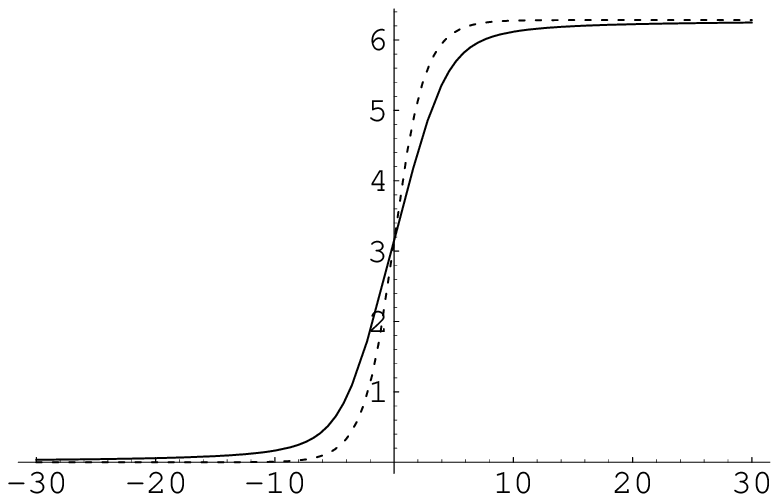}\\
  \caption{Comparison of the standard Yakushevich $n=1$ soliton
\eqref{eq:Ysol}
  (dotted curves) with the $n=1$ solitons (solid curves) obtained
numerically
  by relaxing the $\ell_0 = 0$
  contact approximation (left, from \cite{GaeY1}) and with a Morse
rather than
  harmonic intrapair potential (right, from \cite{GaeY2}). This
clearly shows
  solitons solutions are very little affected by the simplifying
choices used
  in the Y model.}\label{fig:Ycompare}
\end{figure}

It should be mentioned, in this respect, that investigations
conducted within the framework of the simple Y model have shown
that while dispersion relations are of course strongly affected by
the contact approximation and by the choice of the pairing
potential, these have very little effect on the soliton equations
(provided parameters in the pairing potential are set obeying to
the same physical argument and considerations); this is shown in
Figure \ref{fig:Ycompare}.

\section{Soliton's speed selection}
\label{sec:speed}

A  general feature  of sine-Gordon solitons (and more in general
of relativistic solitons) is that the soliton speed $|v|$ is a
free parameter, which can be fixed by choosing initial conditions
and is bounded from above by a limiting value $c_{0}$. This is a
consequence of the Lorentz symmetry of the equation that fixes a
limiting upper bound for the speed, whereas $v$ can be changed by
applying a boost.

Very often it happens that non-linear systems (e.g the DNA chain,
but also reaction-diffusion equations \cite{Mur}, tsunami
equations \cite{Kun}, etc.) that allow for solitary, non
dispersive excitations, somehow select the order of
magnitude for the speed of propagation of this excitation, in the
sense that when solitons are experimentally observed, they turn
out to have a speed of a well defined order of
magnitude.\footnote{We observe that if soliton excitations are
relevant in DNA transcription, they should also have some built-in
speed selection mechanism: in fact, the order of magnitude of the
speed of the transcription bubble along the DNA double helix is
well defined (and such to coordinate with the synthesis of RNA
messenger by RNA Polymerase), as mentioned in the caption to
Figure \ref{fig:trascri}.}

Thus, in practical situations, whenever the experiments give a
well-defined value for the propagation speed of the soliton, the
speed degeneracy represents a loss of predictive power of the
model. It is quite remarkable that our model has a built-in
mechanism for selecting the soliton speed; it is essential for
this mechanism that we have (at least) a two-components system.

\subsection{Speed selection in the composite DNA model}

One can easily realize that the compatibility condition
(\ref{comp}) fixes the speed of propagation of the soliton
(\ref{fe3}) to the speed $c_{t}$ of the transverse sound waves
supported by the elastic torsional forces acting on the disk
\beq\lb{speed} v=c_{t}=\sqrt{\omega_{t}/I}, \feq where
$\omega_{t}=K_{t}\delta^{2}$.

Moreover, as the soliton exist only for $\mu<0$, the soliton speed
is bounded from above by the speed $c_{s}$ of transverse sound
waves supported by the elastic stacking forces acting on the
pendulum, i.e. \beq\lb{speed1}
 v \le c_{s}=\sqrt{\omega_{s}/m}, \feq
 where $\omega_{s}=K_{s}\delta^{2}$.
In view of Eq. (\ref{speed}), this implies a constraint on the
stacking, torsional coupling constants and kinematical parameters
of the system: \beq\lb{h8} \frac{K_{t}}{I}<\frac{K_{s}}{m}. \feq

This selection  mechanism for the soliton speed gives a nice and
simple way to produce solitons with a given speed in double
pendulums molecular chains. To select the soliton speed one just
needs to tune the torsional and stacking coupling constants and
the kinematical parameters of the chain such that Eqs.
(\ref{speed}) and (\ref{h8}) are satisfied. Acting on the
confining potential $U_c$, making it stronger and stronger, one
obtains the single pendulum limit of the double pendulums chain.
The angle is frozen to  $\phi=0$ and a SG soliton with a speed
equal to that of the transverse sound waves supported by the
torsional forces acting on the disk is selected.\footnote{Thus in
a way this mechanism is related to the fact the simple pendulum
model is not structurally stable, and should be seen as the
singular limit of a class of more general systems. Note this class
could be not unique: e.g. for the pendulum case, one could
consider a chain of coupled pendulums made of elastic beams,
obtaining the sine-Gordon equation as the limit case when beams
are infinitely rigid.}

Notice that the mechanism described here can be obviously used to
devise and realize non-linear media where solitons propagate at a
given fixed speed.

\subsection{Speed selection in general models}

The mechanism for selecting the soliton speed described above for
the molecular chain model (\ref{simple}) is rather generic. It is
related to the existence of a conditionally conserved quantity
$J$\footnote{This is the momentum conjugate to the angle $\theta$
evaluated at $\phi=0$ \cite{CDG3}, $J = (\partial \L / \partial
\dot{\theta})_{\phi = 0}$.} and it is rather independent from the
specific form of the interactions characterizing the model. The
proposed mechanism will work whenever we have a nonlinear
mechanical system satisfying some general
conditions:
\begin{enumerate}
\item The system must have at least two degrees of freedom ($X,Y$)
at each site, which in the continuum approximation will give two
interpolating fields ${\hat X}(x,t),\,{\hat Y}(x,t)$ and are
characterized by masses (or moments of inertia)
$m,M$ with $m\neq M$;

\item There should be at least two types of interactions: $(a)$ An
elastic force (coupling constant $K_{t}$) originated by the
interaction between neighboring sites on the chain; $(b)$ A non
linear force (coupling constant $K_{p}$) acting on the single
site;

\item There should be a confining potential $U_{c}$ that limits
the range of variation of one degree of freedom (e.g $Y$) and
allows to freeze $Y$. That is, by making $U_{c}$ steeper and
steeper we implement the dynamical reduction from two degrees of
freedom to one single degree of freedom $X$;

\item Freezing the degree of freedom $Y$ we have both a
conservation law for the momenta conjugate to $X$ and solitonic
solutions for ${\hat X}$.
\end{enumerate}

\section{Perturbative expansion and slaving}
\label{sec:slaving}

A drawback of the general composite Y models (\ref{eq:LAGcYparts})
is that the equations describing its dynamics are far too complex
to be exactly solved.

On the other hand, the doubling of degrees of freedom introduces a
natural separation between topological and non-topological degrees
of freedom. This separation, together with the fact that the
composite model allows for the exact Y soliton (\ref{fe3}) when
the non-topological field $\phi$ is frozen, opens the way to a
perturbative treatment of our model.

The simplest way to deal with our non linear system at the
perturbative level, is to consider it as embedded in a family of
double pendulums chains, which has a single pendulum chain as a
special case.

We will first consider a chain of simple pendulums (of length $R$
and mass $M$); we will then look for solutions of the double
pendulums model by perturbing the system near the single pendulum
Y solutions (\ref{fe3}).

As the simple pendulum limit of our double pendulums chain
involves a reduction of the number of degree of freedom, we will
have to deal with a singular perturbative expansion.

A way to obtain the simple pendulums chain (and the sine-Gordon
equation in the continuum limit) from our model is to let the
length of second pendulums go to zero. In this case our set of
parameters becomes redundant, as the positions of the masses of
the pendulums coincide in space, so that only the total mass is
relevant. A similar argument can be used to show that only the
total coupling strength $\^K = K_s + K_t$ is relevant.

If the double pendulums chain is seen as a (singular) perturbation
of the simple pendulums, one is naturally led to look for
travelling wave solutions as perturbations of the standard
sine-Gordon solitons.

This means we look for solutions to the equations (\ref{eq:TWcY})
in the form of a series expansion in a small parameter $\eps$. We
will correspondingly also expand in the same parameter the
parameters appearing in the model and in the solution: the
geometrical parameters, the masses appearing in our model, the two
coupling constants $K_t$ and $K_s$ and also allow for modification
of the speed by expanding it as well \cite{Gle,Ver}.

Therefore, the series expansion we adopt are as follows:
\beq\label{seriesall}
\begin{array}{lll}
\vth  & = \  \vth_0 + \eps (\de \th)  & = \  \vth_0 + \eps \vth_1 +
\eps^2 \vth_2 + .... \ , \\
\phi  & = \  \eps (\de \phi)  & = \  \eps \phi_1 + \eps^2 \phi_2 +
.... \ ; \\
r  & = \  \eps (\de r)  & = \  \eps r_1 + \eps^2 r_2 + ... \ , \\
R  & = \  A - \eps (\de r)  & = \  A  - \eps r_1 - \eps^2 r_2 - ... \
; \\
m  & = \  \eps (\de m)  & = \  \eps m_1 + \eps^2 m_2 + ... \ , \\
M  & = \  M_{tot} - \eps (\de m)  & = \   M_{tot} - m_0 - \eps m_1 -
\eps^2 m_2 - ... \ ; \\
v  & = \  v_0 + \eps (\de v)  & = \  v_0 + \eps v_1 + \eps^2 v_2 +
... \ ; \\
K_t  & = \  \eps (\de \^K)  & = \  \eps k_1 + \eps^2 k_2 + ... \ , \\
K_s  & = \  \^K - \eps (\de \^K)  & = \  \^K \, - \eps k_1 -
\eps^2 k_2 - ... \ . \end{array} \eeq where $\vth_0$ and $\phi_0$
are the limiting simple pendulum solutions given by (\ref{fe3}).

Inserting the series expansions (\ref{seriesall}) in the equations
for TW solutions (\ref{eq:TWcY}) we will obtain perturbative
solutions of our non linear dynamical system at the various order
of the perturbative expansion in the parameter $\eps$.

We will skip here the computational details, which can be found in
Ref. \cite{CDG2,CDG4}; the results one obtains in this way are
summarized as follows.

\begin{itemize}

\item At zero  order of the perturbation theory the single
pendulum solitonic solution (\ref{fe3}) is reproduced.

\item First and higher order corrections to $\vth_{0}$ and
$\phi_{0}$ can be explicitly calculated. They exhibit the
following striking feature: the non-topological field $\phi_{k}$
turns out to be completely determined algebraically -- we say then
it is {\it slaved} -- order by order by the topological one
$\theta_{k}$. Thus $\vth_{k}$ is obtained as the solution to a
differential equation which depends on $\{\phi_0,...,\phi_{k-1} ;
\vth_0\ldots\vth_{k-1}\}$, while $\phi_k$ is determined
algebraically (no differential equation involved!) by
$\{\phi_0,...,\phi_{k-1} ; \vth_0\ldots\vth_{k-1}\}$.

\item One can expand $\L$ in a series in $\eps$, $\L = \sum_k
\eps^k \L_k$; then at any order in the perturbative expansion the
Lagrangian $\L_k$ depends on $\phi_k$ but it is independent of the
momentum conjugated to $\phi_k$. Thus $\phi_k$ can be considered
as an {\it auxiliary field}, entering in the Lagrangian only
algebraically (and not differentially). We can  express this fact
by saying that the field $\phi$ is a {\it auxiliary field in
perturbation}.
\end{itemize}

\noindent Once again these remarkable features are not specific to
the DNA model considered here, but do quite obviously apply to a
much wider class of mechanical (and field-theoretical) models.

Actually, one can state that whenever a two-component evolutionary
equation can be expanded in series so that one of the two fields
is an auxiliary field in perturbation, then it will be slaved in
the perturbative expansion, and perturbative solutions will admit
the solution of the simpler PDE obtained by freezing the field
which is auxiliary in perturbation to zero as the limit for $\eps
\to 0$.


\section{Solitons in polyethylene crystals}
\label{sec:poly}

In this section we will substantiate our claim that the mechanisms
(in particular, the soliton speed selection and the slaving)
devised in the study of DNA are significant for more general
classes of polymeric macromolecules by showing how they apply to a
Polyethylene (PE in the following) chain in a crystal environment.

The (nonlinear) dynamics of crystalline PE chains has been
discussed in various papers \cite{BV1,GeMan,MaSav,SavMan,SkWo,
Zhang,ZhC0,ZhC} following pioneering modelling work by Kirkwood
\cite{Kirk} and by Mansfield and Boyd \cite{ManBo}; here we will
follow the approach by Zhang and Collins \cite{ZhC}.

\subsection{General features}

It has been argued that the twisting and the elongation (or
compression) of the PE chain can be described by elementary
excitations called {\it twistons}, smooth twists of the PE chain
accompanied with a contraction (or elongation) of the CH$_{2}$
units of the chain \cite{ZhC}. These twiston excitations may be
relevant to describe the propagation of conformational defects in
the PE chain or, more in general, its molecular dynamics. Using
realistic intramolecular and intermolecular interactions, it has
been shown that twisting and elongation of the PE chain can be
described by several types of topological sine-Gordon solitons
\cite{ZhC}. This is essentially due to the symmetry of the
crystals structure, which allows for many equivalent ground states
of the interchain potential. As in the continuum limit the
dynamics of the PE chain is modelled by coupled equations of
(multiple) sine-Gordon type, we expect that the speed selection
discussed in section \ref{sec:speed} and the slaving discussed in
section \ref{sec:slaving} apply also to the case of crystalline PE
chain. Before discussing these effects let us briefly summarize
the relevant features of the Zhang-Collins PE model \cite{ZhC}.

Each CH$_{2}$ unit of the PE chain is considered as a rigid group
with mass $m$; it is labeled by an index $n$ and has both
intermolecular interactions  with the whole crystal environment,
modelled by a potential $U_\a$, and intramolecular interactions
with its next-neighbor CH$_{2}$ units, modelled by a potential
$U_\b$.

Owing to the symmetry of the crystal environment a cylindrical
coordinate system is appropriate. The position of the $n$-th
carbon atom is given by the cylindrical coordinates
$(\rho_{n},\varphi_{n}, \zeta_{n})$.

The dynamics of the PE chain is described by the Lagrangian
\beq\lb{pelag} {\cal L} \ = \ T \ - \ U_\a \ - \ U_\b \ , \eeq
with kinetic energy given by \beq\lb{kin} T \ = \ \frac{m}{2} \
\sum_{n} \( {\dot \rho}_n^2 + \rho^{2}_{n} {\dot \varphi}^2_n +
{\dot \zeta}^2_n \) \ . \eeq The intermolecular potential energy
$U_\a$ is the sum of the effective potential $U$ for each $CH_{2}$
unit \cite{ZhC}: $U_\a = \sum_{n} U(\varphi_{n},\zeta_{n})$. Owing
to the symmetry of the crystal, $U$ has a $\pi$-periodic part
$U_{0}(\varphi_{n} )$ and a $2\pi$-periodic part
$U_{1}(\varphi_{n} , \zeta_{n})$, \beq\lb{potential}
U(\varphi_{n},\zeta_{n}) \ = \ U_{0}(\varphi_{n}) \ + \ B \,
\sin(\varphi_{n}-\bar\varphi_{n}) \, \cos(\pi \zeta_{n}/c) \eeq
where $B,\bar\varphi,c$ are constants and, neglecting a small
constant phase, we have \beq\lb{uo} U_{0} \ = \ - \, A \,
\cos(2\varphi_{n}) \ - \ {\widetilde A} \, \cos(4\varphi_{n}) .
\feq Note that $U (\varphi,\zeta)$ in the whole is thus $2
\pi$-periodic in $\varphi$ and $\pi\zeta/c$.

The intramolecular interaction $U_\b$ can be given in terms of
simple sums of the energy of C-C bonds, C-C-C bond bends and
C-C-C-C torsions \cite{ZhC}.

In order to describe the soliton's speed selection and the slaving
in the PE chain we will consider a simplified model for the PE
dynamics. Note that since the variation of $\rho_n$ is rather
small even in the twisting region \cite{ZhC}, one can treat
$\rho_{n}$ as a constant, at least in a first approximation. We
will therefore set $\rho_{n}=r_{0}$, {\it identically}, in the
Lagrangian (\ref{pelag}). This assumption allows us to get rid of
one degree of freedom, which is not essential for our purposes.

We will also neglect the term in the potential $U_{0}$ which is
$\pi/2$-periodic; that is, we set ${\widetilde A}=0$ in Eq.
\eqref{uo}. This approximation allows us to reduce our Lagrangian
to that describing a set of  coupled double sine-Gordon
equations. As well known, the double sine-Gordon equation is also
integrable \cite{CalDeg} and admits topological soliton solutions,
so that it has to be expected that the general mechanisms at work
in the simple sine-Gordon case would also apply here.\footnote{In
this sense, a rougher approximation reducing the problem to
coupled (simple) sine-Gordon equations would also not change the
qualitative features of the problem, while allowing a considerable
simplification of its analytical treatment. As shown below, albeit
this statement is correct in general, with the physical values of
the PE model we are considering, a certain compatibility condition
applying in the sine-Gordon approximation is not satisfied (and
solitons like those we are interested in cannot exist), while the
equivalent condition is satisfied in the double sine-Gordon model.}

Note that for ${\widetilde A}=0$, the potential \eqref{potential}
has a minimum for $\zeta_{n}=0$ and $\varphi_{n}$ identified by
the condition \beq B \, \cos(\varphi_{n} - \bar\varphi_{n} ) \
\cos(\pi \zeta_{n}/c) \ + \ 2 \, A \ \sin (2\varphi_{n}) \ = \ 0 \
. \eeq In order to keep the calculations as simple as possible we
will determine the phase $\bar\varphi_{n}$ so that $U$ has a
minimum for $\zeta_{n}=0$, $\varphi_{n}=-\pi/2$ (in terms of the
fields $\Psi,\Phi$ introduced in \eqref{fr} below, this
corresponds to $\Psi=\Phi=0$). This condition provides
$\bar\varphi=0$. \footnote{The value of $\bar\varphi_{n}$ given in
Ref. \cite{ZhC}, is slightly different, $\bar\varphi_{n}=0.28
\pi$. Note however that this refers to the general model with
${\widetilde A} \neq 0$, and phases are chosen in \cite{ZhC} so to
have minima of the potential for $\zeta_{n}=\varphi_{n}=0$; thus
we are applying the same criterion used in \cite{ZhC} for the
general model to our simplified case ${\widetilde A} = 0$.}

In the continuum limit $n\to x/c$, where $x$ gives now the
coordinate along the chain axis, the system is described by the
Lagrangian density \beq\lb{laden} {\cal L}= \frac{m}{2}(
r_{0}^{2}\psi_{t}^{2}+u_{t}^{2})-\frac{1}{2}(k_{2}c^{2}\psi_{x}^{2}
+k_{3}c^{2}u_{x}^{2})-B \sin \psi\cos(\frac{\pi u}{c})+ A\cos
2\psi, \eeq where the fields $\psi(x,t),u(x,t)$ represent the
continuum limit of the displacement of the coordinates
$\varphi_{n} (t), \zeta_n (t)$ from their equilibrium values,
$r_{0}$ is the equilibrium value of $\rho$, and the constants $k_2
,k_3$ depend both on the elastic coupling constants of the
intramolecular potential and on the geometrical parameters of the
PE chain \cite{ZhC}.

Using suitable approximations, one can derive twiston solutions
for the Euler-Lagrange field equations stemming from the
Lagrangian (\ref{laden}). These solutions have the form of a kink
for the field $\psi$, topological sine-Gordon soliton for the
field $\pi u/c$ \cite{ZhC}. These
solutions bear a strong resemblance with the solutions of the
composite DNA model described in the previous section. This is not
surprising, as the models themselves are very similar; in
particular, the Zhang-Collins model for polyethylene would
translate into a ``composite'' model of Cocco-Monasson type
\cite{CM} model for DNA.

It is therefore natural to look for a soliton speed selection
mechanism, and a slaving mechanism, similar to those discussed in
section \ref{sec:speed} and in section \ref{sec:slaving}.

\subsection{Soliton's speed selection in the PE chain}

\subsubsection{The double-sine-Gordon model}

Performing in Eq. (\ref{laden}) the field redefinitions
 \beq\lb{fr} \psi=\Psi-\frac{\pi}{2},\quad
u= \frac{c}{\pi}\Phi, \eeq
the Lagrangian  becomes \beq\lb{laden1} {\cal L}= \frac{m}{2}(
r_{0}^{2}\Psi_{t}^{2}+\frac{c^{2}}{\pi^{2} }\Phi_{t}^{2})-
\frac{1}{2}(k_{2}c^{2}\Psi_{x}^{2} +k_{3}\frac{c^{4}}{\pi^{2}}
\Phi_{x}^{2})+B\cos\Psi\cos \Phi - A\cos 2\Psi \ . \eeq Notice
that now both $\Psi$ and $\Phi$ can be considered as angular
coordinates with $0\le \Psi,\Phi\le 2\pi$. Introducing the
symmetric and antisymmetric field combinations \beq\lb{sa} \chi_+
\ = \ (\Psi+\Phi) \ , \ \ \chi_- \ = \ (\Psi-\Phi) \ , \eeq the
field equations for travelling waves solutions $\chi_+(\xi)$,
$\chi_-(\xi)$ (where $\xi = x\pm vt$) give \beq\lb{fe}
\begin{array}{l}
\mu \, {\chi_+}'' \ + \ J \, {\chi_-}'' \ + \ 2 \, B \, \pi^2 \,
\sin \chi_+ - 4A\pi^2  \sin(\chi_{+}+\chi_{-})\ = \ 0 \ , \\
\mu \, {\chi_-}'' \ + \ J \, {\chi_+}'' \ + \  2 \, B \, \pi^2
\, \sin \chi_- - 4A\pi^2  \sin(\chi_{+}+\chi_{-})\ \ = \ 0 \ ,
\end{array} \eeq
where the prime denotes derivation with respect
to $\xi$ and $\mu,J$ are given in terms of the elastic and
geometric parameters of the model by
\beq\lb{x1}
\begin{array}{l}
\mu \ = \ m \, v^2 \, (\pi^2 r_0^2 + c^2 ) \ - \ c^2 \, (k_2 \pi^2
+ k_3 c^{2}¥) \ , \\
J \ = \ m \, v^2 \, (\pi^2 r_0^2 - c^2 ) \ - \ c^2 \, (k_2
\pi^2 - k_3 c^{2}¥) \ . \end{array} \eeq

The system (\ref{fe}) gives a nice and simple description of the
propagation of {\it  twistons} in terms of symmetric and
antisymmetric combinations of twisting and elongation modes in the
PE chain.  It has the form of two coupled equations of sine-Gordon
type.

The system has many degenerate ground states $\chi_-= 2 n_1 \pi$,
$ \chi_+=2 n_2 \pi$, with $n_1, n_2$ integers. We therefore expect
the existence of  topological sine-Gordon solitons connecting all
these vacua. We will not discuss here the general solutions of the
system \eqref{fe}, but just point out that the system gives a very
simple realization of the speed selection mechanism discussed in
Sect. \ref{sec:speed}.

If we are interested in the symmetric solutions of the system
 we can set in Eqs. (\ref{fe}) $\chi_- = 0$. This can be realized
 dynamically
 introducing in the Lagrangian (\ref{laden1}) a confining
potential
$U_{c}(\chi_-)$ satisfying $(dU_{c}/d\chi_-)_{0}=0$ and rising
(sharply) with $|\chi_-|$, so to freeze the degree of freedom
$\chi_-$ to its vacuum configuration $\chi_-=0$ (i.e. it forces
symmetric configurations, $\Phi=\Psi$). It is easily seen that for
$\chi_-=0$ the field equations
\eqref{fe} become
\beq\lb{sol} \mu
\, {\chi_+}'' \ + (\ 2 \, B -4 A)\, \pi^2 \, \sin \chi_+ \ = \ 0 \ ,
\eeq
whereas the compatibility condition between the two equations
\eqref{fe} reads
\beq\lb{comp1} J\,B+ 2A(\mu -J) \ = \ 0 \ . \eeq
Comparing Eq.
\eqref{sol} with Eq. \eqref{ge1} one easily realizes that for
$\mu/(B-2A)<0$, Eq. \eqref{sol} admits  SG solitonic solutions given by
\eqref{fe3}. The compatibility condition \eqref{comp1} fixes the
speed of the soliton
\beq\lb{fs} v \ = \ V^{s}_{(+)} \ = \ \pm \,
c \ \sqrt{\frac{k_{2}\pi^{2}\omega- k_{3}c^{2} }{
m(\pi^{2}r_{0}^{2}\omega-c^{2})}} \ , \eeq
where $\omega$  is a dimensionless parameter given by
\beq\lb{omega}
\omega=\frac{B}{B-4A}.
\feq
For the values of parameters $A,B$ proposed by Zhang and Collins
\cite{ZhC} and reported in Table 3 we have $\omega=6.33$.
Moreover, from these  values of the parameters $A,B$ it follows
$B> 2A$, so that the condition for the existence of solitonic solutions
becomes $\mu<0$.

\begin{table}
\label{table:cY5}
\begin{center}
   \begin{tabular}{||ccl||ccl||}
    \hline
    $m$ &=& 14.1 \, g/mol & $k_2$ &=& 109.263 \, kJ/mol \\
    $r_0$ &=& 0.4236 \, \AA & $k_3$ &=& 1860.655 \, kJ/(\AA$^2$ mol) \\
    $c$ &=& 1.274 \, \AA & $B$ &=& 1.52 \, kJ/mol \\
    $A$ &=& 0.32\,  kJ/mol & $\omega$ &=&6.33\\
    \hline
\end{tabular}
\caption{The dynamical, geometrical and coupling parameters in the
Zhang-Collins polyethylene model, as given in \cite{ZhC}.}
\end{center}
\end{table}

This condition
implies a maximum speed for the soliton \beq\lb{fs1} |v| \ < \ V_M
\ = \ c \ \sqrt{\frac{ k_{3}c^{2} +k_{2}\pi^{2}}{
m(c^{2}+\pi^{2}r_{0}^{2})}} \ . \eeq

Needless to say, forcing $\chi_+ = 0$ we would obtain a fixing of
the soliton solution speed for antisymmetric fields provided by
the same compatibility condition (\ref{comp1}), so that Eqs.
(\ref {fs}) and (\ref{fs1}) still hold for the antisymmetric solution.

Note that the argument of the square root in \eqref{fs} could be
negative, depending on the values of the parameters.  When this
happens, no symmetric or antisymmetric travelling wave solution is
possible. Moreover, the  soliton speed (\ref{fs}) has to be
smaller than the maximum allowed speed (\ref{fs1}).

Let us now check the existence of the symmetric and antisymmetric
soliton solutions using realistic values for the physical parameters
of the model.
The values of the physical parameters characterizing the PE chain
\cite{ZhC} are collected in Table 3; they indicate that
$k_{2}\pi^{2}\omega$ is much bigger (at least two times) than
$k^{3}c^{2}$,
whereas $\pi r_{0}\omega > c$, allowing for both symmetric and
antisymmetric solitons with fixed speed
\beq\lb{fsn}
V^{s}_{(\pm)}\simeq 7 \times 10^{3}m/s.
\feq

On the other hand the maximum soliton speed (\ref{fs1}) turns out to
be well above the speed of the symmetric or antisymmetric solutions.
From the values of the parameters of Table 3 we get
\beq\lb{ms}
V_{M}\simeq 1.2\times 10^{4} m/s.
\feq

\subsubsection{The sine-Gordon approximation}

We would like to stress that the soliton's speed selection
mechanism is a general and rather robust effect, which does not
depend on the details of the model we are considering -- but the
very existence of soliton may depend on the value of physical
parameters. To have a flavor of this fact let us consider (as
anticipated in a footnote above) an even more simplified model for
the PE.

We will neglect completely the $\pi$-periodic part in the
potential $U(\psi,u)$ in \eqref{potential}, i.e. we will set there
$U_{0}=0$; note this does not change the overall periodicity
properties of $U_\a$. This is a very rough approximation, but it
allows us to reduce our Lagrangian (\ref{laden}) to that
describing a simple two-fields sine-Gordon model. This  situation
represents just the particular case $A=0$ ($\omega=1$) of our
general equations.

By forcing  the degree of freedom $\chi_-$ to its vacuum
configuration $\chi_-=0$ (the same discussion would apply
interchanging the roles of $\chi_+$ and $\chi_-$), the field
equations (\ref{sol})
 become now
\beq\lb{sol4} \mu \, {\chi_+}'' \ + \ 2 \, B \, \pi^2 \, \sin
\chi_+ \ = \ 0 \ , \eeq whereas the compatibility condition
(\ref{comp1}) simplifies to $ J  =  0 $. This selects the soliton
speed,
 \beq\lb{fsm} v \ = \ V^{s}_{(\pm)} \ = \ \pm \, c \
\sqrt{\frac{ k_{3}c^{2} - k_{2}\pi^{2}}{
m(c^{2}-\pi^{2}r_{0}^{2})}} \ . \eeq
 The values of the physical
parameters of Table 3 indicate that $k_{3}c^{2}$ is much bigger
(at least three times) than $k_{2}\pi^{2}$. Equation \eqref{fsm}
requires therefore $c\ge \pi r_{0}$.

The condition for the existence of solitonic solutions (\ref{fs1})
together with the previous equation  imply that the soliton
solutions of our simplified model can exist only for $r_0^2 <
(k_{2}/k_{3})$. The values of the parameters reported in Table 3
do {\it not } satisfy this condition.  Thus for the values of the
parameters proposed by Zhang and Collins, the symmetric and
antisymmetric soliton solutions \eqref{sol} of the more general
model \eqref{laden} are allowed, but the rougher approximation
obtained setting $A=0$ is not allowing such solutions.

\subsection{Speed selection and conditionally conserved quantities}

In Ref. \cite{CDG3} it has been pointed out that the speed
selection mechanism is related to the existence of a conditionally
conserved quantity. It is not difficult to identify this quantity
for the PE model we are discussing here. For the case in which a
confining potential for the field $\chi_-$ is introduced, from the
field equations (\ref{fe}) one can easily derive the following
equation \beq\lb{s1} \frac{d W_+} { d \xi} \ = \ F_+ \ , \eeq
where
\beq\lb{k1}
\begin{array}{rl}
W_{+} = &\left(B J+2A(\mu-J)\right) \chi_{+}'+\left(B
\mu+2A(\mu-J) \right)\chi_{-}',\\
F_+ = & - 2\pi^{2} B \left[ (2A-B)\sin (\chi_-) + 4 AB
\left(\sin(\chi_{+}+\chi_{-})-\sin\chi_{+}\right)\right]\\
 & - [dU_{c}(\chi_-)/d\chi_-].
\end{array}
\eeq If $\chi_-=0$, then $F_{+}¥(\chi_-=0)=0$ and $W_{+} $ is
conserved. Moreover, when $\chi_-=0$, the conserved quantity
becomes $W_{+} =(BJ +2A(\mu-J)) \chi_+'$, which is compatible
with the existence of solitonic solutions to \eqref{sol} only for
$BJ +2A(\mu-J)=0$, i.e. only when the soliton speed is given by
\eqref{fs}.

Similarly, when  a confining potential $U_{c}(\chi_+)$ for the
field $\chi_+$ is introduced, we have the conditionally conserved
quantity $W_{-}$ and a force $F_-$, whose expressions are obtained
by  interchanging $\chi_{+}$ and $\chi_{-}$ in \eqref{k1}. Again,
when $\chi_+$ is frozen to its vacuum value $\chi_+=0$, $W_{-}$ is
conserved and the speed of the $\chi_-$-soliton (if this is
allowed) is fixed to the value \eqref{fsm}.

\subsection{Series expansion, and slaving}
\label{PEslaving}

\def\Pd{{P'}}
\def\Pdd{{P''}}
\def\Qd{{ Q'}}
\def\Qdd{{Q''}}

In order to discuss series expansion around a given solution, we
will use the coordinates $\chi_\pm$, and for notational
convenience we rewrite these as \beq P  \ = \ \chi_+ \ , \ \ Q \ =
\ \chi_- \ ; \eeq moreover, we will work directly in the space of
functions of the variable $\xi = (x \pm v t)$.

\subsubsection{The double-sine-Gordon model}

In this notation, and denoting again $\xi$ derivatives by a prime,
the Lagrangian \eqref{laden1} can be written as \beq\label{LsimpA}
\L \ = \ \frac{1}{2} \left[ M (\Pd^2 + \Qd^2 + 2 S \Pd \Qd )
\right] \, + \, 2 B ( \cos P + \cos Q ) \ - \ 4 A \, \cos (P+Q) \
, \eeq  where we have further simplified the notation by writing
\beq\label{LsimpMS} M = \frac{\mu}{\pi^2} \ , \ \ S =
\frac{J}{\pi^2} \ . \eeq

We will consider a Lagrangian $\L_\eps$ depending on a parameter
$\eps$ so that for $\eps \to 0 $ it reduces to a (sine-Gordon)
Lagrangian $\L_0$ which only depends on $(P,\Pd)$ plus a term
constraining $Q=0$, while for $\eps = 1$ it is just the Lagrangian
\eqref{LsimpA}. The simplest Lagrangian with these properties is
 \beq\label{Leps} \L_\eps \ = \ \frac{1}{2} \left[ M
 (\Pd^2 + \eps \Qd^2 ) + 2 S \eps \Pd \Qd  \right] \, + \, 2 B
( \cos P + \cos Q ) \, - \, 4 A \eps \cos (P+Q)  \ . \eeq
 The corresponding Euler-Lagrange equations are
 \beq \begin{array}{l}
 M \, \Pdd \ + \ \eps \, S \, \Qdd \ + \ 2 \, B \, \sin (P) \ + \
 4 \, A \, \eps \, \sin (P+Q) \ = \ 0 \ , \\
 \eps \, M \, \Qdd \ + \ \eps \, S \, \Pdd \ + \ 2 \, B \, \sin (Q) \ + \
 4 \, A \, \eps \, \sin (P+Q) \ = \ 0 \ .
 \end{array} \eeq

We will then consider $\eps$ series expansions both for the
functions of $\xi$ and for the parameters; thus we write
\beq\label{series}
\begin{array}{rl}
P =& P_0  + \eps P_1  + \eps^2 P_2  + \eps^3 P_3 + ... \ , \\
Q =& Q_0  + \eps Q_1  + \eps^2 Q_2  + \eps^3 Q_3 + ... \ ; \\
A =& A_0 + \eps A_1 + \eps^2 A_2 + \eps^3 A_3 + ... \ , \\
B =& B_0 + \eps B_1 + \eps^2 B_2 + \eps^3 B_3 + ... \ ; \\
M =& M_0 + \eps M_1 + \eps^2 M_2 + \eps^3 M_3 + ... \ , \\
S =& S_0 + \eps S_1 + \eps^2 S_2 + \eps^3 S_3 + ... \ .
\end{array} \eeq
 Note we are also expanding in series the constants $B,M,S$, as prescribed
by the general Poincar\'e-Lindestedt procedure.\footnote{Moreover,
$M$ and $S$ depend not only on the couplings and the geometric
parameters of the model, but also on the wave speed $v$, which
should in any case be allowed to vary with $\eps$.  It may be
worth remarking that the terms of the parameter expansions would
be determined by imposing the projection of higher order terms in
the expansion for $P$ on the space of solutions to the sine-Gordon
equation to vanish. As we are not going to discuss this aspect
(that is, implicitly, how the speed of the solution depends on
$\eps$), but only in displaying slaving of the $\Phi$ field --
i.e. of the $Q$ variable -- these terms will remain undetermined
here. The reader can easily check that slaving shows up as well by
just setting $(A,B,M,S)=(A_0,B_0,M_0,S_0)$.}

We now expand $\L_\eps$ in $\eps$, \beq \L_\eps \ = \ L_0 + \eps
L_1 + \eps^2 L_2 + \eps^3 L_3 + ... \ . \eeq

At order zero, we get (as required) \beq\label{lag0} L_0 \ = \
\frac{1}{2} \ M_0 \Pd_0^2 \ + \ 2 B_0 \, \cos (P_0) \ + \ 2 B_0 \,
\cos (Q_0) , \eeq which yields the sine-Gordon equation for $P_0$,
\beq\label{sgP0} \Pdd_0 \ = \ - \ (2 B_0/M_0) \ \sin (P_0) \ ,
\eeq together with $\sin (Q_0 ) = 0$; this in turn enforces (up to
a suitable choice of the origin for $\Phi$) \beq Q_0 \ = \ 0 \ .
\eeq

At order $\eps$, we get \beq L_1 \ = \ \frac{1}{2} \ M_1 \Pd_0^2 \
+ \ M_0 \Pd_0 \Pd_1 + 2 B_1 [1 + \cos (P_0)] - 2 B_0 P_1 \sin
(P_0) \ + \ 4 A_0 \cos (P_0) \ ; \eeq the Euler-Lagrange equation
with respect to $P_1$ is identically satisfied (it just yields
back the sine-Gordon equation for $P_0$).

At order $\eps^2$, we get \beq \begin{array}{rl} L_2 \ =& \ (1/2)
[ M_0 \Pd_1^2 + M_2 \Pd_0^2 ] \ + \ M_1 \Pd_0 \Pd_1 \ + \
\Pd_0 (M_0 \Pd_2 + S_0 \Qd_1) \ + \\
 & \ - \ B_0 Q_1^2 \ + \ (2 B_2 - B_0 P_1^2 + 4 A_1) \, \cos (P_0) \ + \\
 & \ - \ 2 [B_1 P_1 + B_0 P_2 + 2 A_0 (P_1 + Q_1)] \, \sin(P_0)  \ .
\end{array} \eeq

The equation issued by variation with respect to $P_1$, taking
into account that $P_0$ solves \eqref{sgP0}, reads \beq
\label{eqP1} M_0 \, \Pdd_1 \ = \ - \ 2 [B_0 M_0 \cos (P_0 )]  \
P_1 \ + \ (2/M_0) [(B_0 M_1 - B_1 M_0 ) - 2 A_0 M_0] \, \sin (P_0
) \ . \eeq

As for the variation with respect to $Q_1$ (obviously taking again
\eqref{sgP0} into account), it yields \beq Q_1 \ = \ \(
\frac{S_0}{M_0} \, - \, 2 \, \frac{A_0}{B_0} \) \ \sin (P_0) \ .
\eeq This is just a relation between $Q_1$ and $P_0$, and uniquely
determines $Q_1$ in terms of the latter and of the parameters
appearing in the equation. In particular, it is {\it not } a
differential equation, and shows that $Q$ (that is, the field
$\Phi$ in the notation used earlier on) is slaved to $P$ (that is,
to the field $\Psi$) at leading order.

Actually, it is easy to see that the same will happen at all
orders and not just at leading order. That is, the terms $P_k$
will be determined by linear non-autonomous differential
equations, the non-autonomous terms of these depending on terms
$\{ P_0 , P_1 , ... , P_{k-1} ; Q_1 , ... , Q_{k-1} \}$ of lower
degree; on the other hand, the terms $Q_k$ will be determined {\it
algebraically } in terms of the $\{ P_0 , ... , P_k
\}$.\footnote{Using this fact, the equations for the $P_k$ can be
written in terms of the $P_j$ (with $j < k$) alone, i.e. without
explicit use of the $Q_j$.}

Needless to say, the explicit equations (and the partial
Lagrangians $L_k$ too) will become quickly quite involved. Thus we
will just give the equations satisfied by $\eps^2$ order terms for
$P$ and $Q$. The term $P_2$ is obtained as solution to the
differential equation \beq
\begin{array}{rl}
 B_0 M_0^3 \Pdd_2 =&  [(B_0 M_0 S_0^2 - 2
A_0 M_0^2 S_0) \sin (P_0)] \ \Pd_0^2 \ + \ [B_0 M_0 \cos (P_0)] \,
P_2 \ + \\
& - \ 2 [B_0 M_0 (2 A_0 M_0 + B_1 M_0 - B_0 M_1) \cos(P_0)] \ P_1
\ + \\
& + [B_0^2 M_0^2 \sin (P_0) ] P_1^2 +
 2 [ (2 A_0 + B_1) B_0 M_0 M_1 - (2 A_1 + B_2) B_0 M_0^2 + \\
 & \ + (M_0 M_2 - M_1^2) B_0^2 + (B_0 S_0 -2 A_0 M_0 )^2 \cos(P_0)]
\, \sin(P_0) \ ; \end{array} \eeq as for $Q_2$, this is given by
\beq
\begin{array}{rl}
 Q_2  =&  (2 B_0^2 M_0^2 )^{-1} \
 \left[ \( 2 A_0 M_0 (2 (B_1 M_0 + B_0 S_0) - M_0^2 \Pd_0^2) +
 \right. \right. \\
 & \left. \left. \ + \ B_0 (2 B_0 (M_0 S_1 - M_1 S_0)
  - 4 A_1 M_0^2 + M_0^2 S_0 \Pd_0^2) \) \, \sin(P_0) \ + \right.
  \\
 & \left. \ + \ 2 M_0 (2 A_0 M_0 - B_0 S_0) \cos (P_0)
 ((2 A_0 - B_0) \sin (P_0) - B_0 P_1 ) \right] \ .
 \end{array} \eeq

\subsubsection{The sine-Gordon approximation}

Slaving of the $Q$ degree of freedom is present also in the
sine-Gordon approximation considered above, i.e. for $A=0$. In
this case one obtains slightly simpler explicit formulas, easily
obtained by setting $A_0=A_1 = 0$ in the previous ones.

As recalled above, see the footnote following Eq.\eqref{series},
if one is only interested in slaving a simpler series expansion --
with $A,B,M,S$ not depending on $\eps$ -- would also be possible.
We will now adopt this simplified expansion with the sine-Gordon
approximation (i.e. $A=0$) in order to see the slaving mechanism
at work for the PE model in its simplest setting.

With the same notation as above, the Lagrangian \eqref{laden1}
with $A=0$ reads \beq\label{Lsimp} \L \ = \ \frac{1}{2} \left[ M
(\Pd^2 + \Qd^2 + 2 S \Pd \Qd ) \, + \, 2 B ( \cos P + \cos Q )
\right] \ . \eeq The Lagrangian $\L_\eps$ can be chosen as
\beq\label{LepsSG} \L_\eps \ = \ \frac{1}{2} \left[ M
 (\Pd^2 + \eps \Qd^2 ) + 2 S \eps \Pd \Qd \, + \, 2 B
( \cos P + \cos Q ) \right]  \ , \eeq and the corresponding
Euler-Lagrange equations are
 \beq \begin{array}{l}
 M \, \Pdd \ + \ \eps \, S \, \Qdd \ + \ 2 \, B \, \sin (P) \ = \ 0 \ , \\
 \eps \, M \, \Qdd \ + \ \eps \, S \, \Pdd \ + \ 2 \, B \, \sin (Q) \ = \ 0 \ .
 \end{array} \eeq

With the simplified series expansions, and expanding $\L_\eps$ in
$\eps$ as before, we have partial Lagrangians $L_k$ corresponding
to terms of order $\eps^k$.

At order zero, we get of course \eqref{lag0} (except for $B_0,M_0$
now reading $B,M$), which yields the sine-Gordon equation for
$P_0$, \beq\label{sgsimp} \Pdd_0 \ = \ - \ (2 B/M) \ \sin (P_0) \
, \eeq together with $Q_0=0$.

At order $\eps$, we get \beq L_1 \ = \ M \Pd_0 \Pd_1 \ - \ 2 B P_1
\sin (P_0) \ , \eeq with the Euler-Lagrange equation with respect
to $P_1$ identically satisfied (it just yields back the
sine-Gordon equation for $P_0$).

At order $\eps^2$, we have \beq L_2 \ = \ (1/2) ( M \Pd_1^2 )  +
\Pd_0 (M \Pd_2 + S \Qd_1)  -  B Q_1^2 \ . \eeq The equation issued
by variation with respect to $P_1$ yields, taking into account
that $P_0$ solves \eqref{sgsimp}, \beq \label{eqP1} M \, \Pdd_1 \
= \ - \ 2 \( B M \cos (P_0 ) \) \ P_1 \ . \eeq

As for the variation with respect to $Q_1$, taking again
\eqref{sgsimp} into account it yields \beq M \ Q_1 \ = \ 2 B S
 \, \sin (P_0) \ . \eeq Thus again we get a relation between
$Q_1$ and $P_0$ (not a differential equation), which uniquely
determines $Q_1$ in terms of the latter and of the parameters
appearing in the equation.

Albeit simpler than in the general setting, explicit expressions
still become quite involved at higher orders, and we just give the
equations satisfied by $\eps^2$ terms. As for $P_2$, it solves the
equation \beq M^3 \Pdd_2 =  - 2 B M^2 \cos (P_0)  P_2  + [B M^2
P_1^2 +  M S^2 \Pd_0^2 + 2 B S^2 \cos (P_0) ] \sin (P_0) \ .
 \eeq
The next-to-leading order term $Q_2$ for $Q$ is given by \beq
 Q_2  =  (2 B M^2 )^{-1} \ [ M^2 S \Pd_0^2 \ + \
 2 B M S \cos (P_0) (P_1 + \sin (P_0)) ] \ . \eeq

\section{Conclusions and discussion}
\label{sec:concl}

Almost thirty years after the seminal paper by Englander,
Kallenbach, Heeger, Krumhansl and Litwin, non linear mechanical
models of DNA still represent an active area of research, and a
toll for trying to tackle fundamental problems such as the
denaturation and transcription processes.

The nice feature of this mechanical approach, not shared by
approaches using full molecular dynamics, lies in its simplicity.
This simplicity allows to model general features of DNA and to
extract relevant information with relatively few analytical and/or
computational effort.

Simple DNA mechanical models are obviously too simple to take into
account the full complexity of the DNA macromolecule; on the other
hand, they may well be able to describe DNA dynamics for what
pertains to specific biological processes -- such as DNA thermal
denaturation or the formation and dynamics of open bubbles to
which RNA Polymerase could bind in the transcription process.

Simple DNA mechanical models as the ones formulated by Peyrard and
Bishop and by Yakushevich were able to provide correct qualitative
predictions, and fit the order of magnitude of biologically
relevant and physically observable quantities (e.g. the frequency
of small amplitude oscillations, characteristic scales of
breathers and some statistical mechanics features
\cite{GRPD,PeyNLN,PBD,PD} in the denaturation transition, and the
size of solitonic excitations \cite{GaeJBP,GRPD,PD,YakuBook} in
the context of DNA transcription); on the other hand they failed
completely when confronted to other physical quantities which are
directly observable in modern single-molecule experiments
\cite{Lavery,Ritort} and which involve elastic properties of the
DNA molecule, such as the speed of transverse elastic waves
\cite{YakPRE}.

The new generation of DNA mechanical models, in which there are
more than one degrees of freedom per nucleotide, such as the BCP
and CM models in the context of DNA denaturation
\cite{BCP,BCPR,CM,PeyNLN,PB}, and the composite model discussed in
this note for what concerns torsional DNA dynamics and
transcription, represent a big improvement in the direction of a
more accurate modelling of DNA still retaining the attractive
features of simple models.

In fact, on the one hand they remain simple enough so that their
dynamics can be at least controlled, if not completely solved, at
analytical level; on the other hand they allow for a more
realistic description of the DNA complexity.

Focusing on torsional DNA dynamics and transcription, if one
considers our composite model then the predicted speed for optical
and sound excitation in the DNA chain fits the order of magnitude
of the experimental data that cannot be fitted by the simple Y
model with physically acceptable coupling constants.

Moreover, the greater number of degrees of freedom per site -- and
more specifically the fact one of these refers to the homogeneous
(sugar-phosphate backbone) of the DNA molecule, the other to the
non-homogeneous (nitrogen Watson-Crick bases) of it -- enables one
to introduce in a natural way those inhomogeneities in the DNA
chain (in the form of different basis sequences) that are
necessary for the codification of the genetic information.

We also stress that real DNA lives in a highly viscous (at the
molecular scale) fluid and is subject to thermal noise. These
features could be implemented more realistically within a
composite model able to take into account the differences between
the external (backbone) and the internal (bases) parts of the DNA
molecule.

Apart from the specific problems of DNA modelling, and maybe more
relevantly for the general community working in nonlinear systems
and nonlinear Mechanics, the research activity on DNA dynamical
modelling has also contributed to focusing on previously unnoticed
mechanisms and deepen our understanding on nonlinear phenomena; it
thus became a source of new ideas in the field.\footnote{For a
discussion of this statement in relation to breather-type
nonlinear excitations, we refer to the book of Peyrard and Dauxois
\cite{PD}.}

In a separate but related development, Saccomandi and Sgura
\cite{SacSgu} have realized that chains with fully nonlinear
elastic nearest neighbor coupling would present peculiar features,
in particular in these the solitonic excitations would have a
strictly finite size -- thus be {\it compactons} rather than
ordinary solitons (see also \cite{Takcomp} in this respect). This
mechanism can also be generalized and extended to more general
systems than the DNA molecule \cite{DGS,GaeEPL,GGW}.

For what concerns the model discussed here \cite{CDG1,CDG2}, the
mechanism for selecting the speed of solitons by tuning the
physical parameters of the system on the one hand \cite{CDG3}, and
the separation in slaving and master fields \cite{CDG4} described
in this note are two nice examples of this statement.

In particular, the speed selection mechanism which has been
originally discovered for the DNA composite model
\cite{CDG1,CDG3}, can be generalized for an ample class of
molecular chain models \cite{CDG3} and could find broad
applications to devise and realize nonlinear media where solitary
wave excitations propagate at a selected speed.

Also the separation of the  degrees of freedom in ``master'' and
``slave'' seems not to be limited to DNA non linear dynamics, but
to be a quite generic feature of this ample class of nonlinear
systems. It may be very useful for separating in a hierarchical
way the different degrees of freedom  that are relevant to the
dynamics of the nonlinear system, and be a guiding principle to
make easier perturbation analysis of such systems.

We illustrated the possibility of a wider applications of these
two mechanism in the study of (polymeric) macromolecules by
considering a concrete different system, i.e. polyethylene, and
showing by explicit computations how they apply in this case as
well.

\subsection*{Acknowledgements}

We thank M. Barbi, S. Cuenda, T. Gramchev, M. Joyeux, G.
Saccomandi, A. Sanchez, I. Sgura and S. Walcher for useful
discussions on DNA dynamics and related matters over the last few
years. We would also like to thank the Editor for urging us to
work out the extension of previous results given in section
\ref{sec:poly}.

\end{document}